# Self-focusing and the Talbot effect in conformal transformation optics

Xiangyang Wang[1,a], Huanyang Chen[2,a,b], Hui Liu[1,b], Lin Xu[2], Chong Sheng[1] and Shining Zhu[1]

[1]National Laboratory of Solid State Microstructures & School of Physics, Collaborative Innovation Center of Advanced Microstructures, Nanjing University, Nanjing, Jiangsu 210093, China.
[2]Institute of Electromagnetics and Acoustics and Department of Electronic Science, Xiamen University, Xiamen 361005, China.
[a]These authors contributed equally to this work.
[b] Corresponding author：e-mail: kenyon@xmu.edu.cn; liuhui@nju.edu.cn.

**Abstract:** Transformation optics (TO) has been used to propose various novel optical devices. With the help of metamaterials, several intriguing designs, such as invisibility cloaks, have been implemented. However, as the basic units should be much smaller than the working wavelengths to achieve the effective material parameters, and the sizes of devices should be much larger than the wavelengths of illumination to work within the light-ray approximation, it is a big challenge to implement an experimental system that works simultaneously for both geometric optics and wave optics. In this letter, by using a gradient-index micro-structured optical waveguide, we realize a device of conformal transformation optics (CTO) and demonstrate its self-focusing property for geometry optics and Talbot effect for wave optics. In addition, the Talbot effect in such a system has a potential application to transfer digital information without diffraction. Our findings demonstrate the photon controlling ability of CTO in a feasible experiment system.

In gravitational lensing[1], light rays are bent by a star because its gravitation changes the geometric property of space. Similarly, the propagation direction of light is altered in a transmitting medium relative to a vacuum because of the interaction between the electromagnetic field and matter[2]. Based on an analogy between spacetime geometry and light propagation in a medium, two papers on invisibility cloaks[3,4] started research into transformation optics (TO)[5-7], which deepened our understanding of gravitational analogues in optical systems. Furthermore, general relativity in electrical engineering (an analogue electromagnetic system) has been proposed[8-10] and implemented at visible frequencies[11-14].

In the last decade, developments in material science have enhanced our ability to design optical devices[6] and control electromagnetic waves using TO. Moreover, the principles of transformation optics can be harnessed to control other kinds of wave[15-18]. Although TO is a very beautiful theory and has many fantastic applications, it has encountered some difficulties in experiments. In the earlier work on invisibility cloaking[3,19], different kinds of split-ring resonance structures are used to tune the effective material parameters, which are inhomogeneous and anisotropic tensors. The working wavelengths should be much larger than the resonance units. On the other hand, in order to satisfy the light-ray approximation, the working wavelengths should be much smaller than the sizes of devices. For this reason, it is a great challenge to experimentally realize both geometric optics and wave optics in a single transformation-optical device[20]. In geometric optics, light is treated as particles, and the only important thing is the trajectories. But in wave optics not only do the trajectories matter, but also phase changes play a very important role.

In two-dimensional space, conformal transformation optics (CTO)[4,21], as a branch of TO, can steer light rays by using a dielectric medium with an inhomogeneous and isotropic refractive index profile. Light rays (in the geometric-optics regime) can be bent by such a dielectric medium. Recently, conformal transformation optics was further expanded into the realm of wave optics, and many groups have done extensive studies in this field, such as cloaking[22-26], whispering gallery modes[27], broadband plasmonic devices[28,29], Casimir effect[30] and analysis of electron energy loss[31]. So devices from CTO might be good candidates for working simultaneously in the geometric-optics and wave-optics regimes.

In this letter we introduce a conformal lens, also known as the Mikaelian lens[32]. This conformal lens is mapped from the Maxwell's fish-eye lens by an exponential conformal mapping[33]. We construct such a conformal lens with a gradient-index micro-structured optical waveguide and observe that it can self-focus a beam: a geometric-optics property. Moreover, we see the Talbot effect in the same lens, which stems from a phase change and only happens for wave optics.

The Talbot effect was firstly discovered in 1836[34] and was explained by Rayleigh in 1881[35]. It was rediscovered at the beginning of the 20th century[36-38], and in the mid-1950s Cowley and Moodie revisited this effect[39,40], which received much attention[41-44]. Recently the Talbot effect has been realized in several different systems, such as metamaterials[45] and surface plasmonics[46], which have many applications[47,48]. By proper design, we find that the Talbot effect in the conformal lens can be further applied to transfer digital information without diffraction. We verified these conformal Talbot effects through experiment measurements, numerical simulation and analytic calculations.

Let us first recall the basic principle of CTO[4,21]. In a two-dimensional space denoted by $w = u + vi$, if there is a refractive index profile $n_w(u+vi)$, light rays will propagate along curved trajectories if $n_w$ is not uniform. Considering another space denoted by $z = x + yi$, related by a conformal mapping ($w = w(z)$) which satisfies Cauchy-Riemann condition[49], one can construct a point-to-point corresponding relationship between $u-v$ space and $x-y$ space. If the refractive index profile $n_w(u,v)$ in $u-v$ space and $n_z(x,y)$ in $x-y$ space satisfy[4]

$$n_z = \left|\frac{dw}{dz}\right| n_w, \tag{1}$$

then light trajectories in $x-y$ space can be simply obtained by conformal mapping from those in $u-v$ space according to CTO (see Supplemental Material Section I for more details[50]). Therefore Eq. (1) establishes the corresponding relationship of light propagation between two spaces by conformal mapping.

Now we introduce a CTO device that starts from the well-known Maxwell's fish-eye lens in two dimensions. Its refractive index profile is $n_w(u,v) = 2\alpha/(1+u^2+v^2)$, where $2\alpha$ is the refractive index at the centre (Fig. 1(a)). By using the variational method to obtain light trajectories[54], we know that all the light rays emitted from a point source at point **A** will travel along the solid red circles and converge to an image at point **B.** Suppose we have an exponential conformal mapping $w = \exp(\beta z)$, which can map a $u-v$ complex plane in Fig. 1(a) to a ribbon-like region of $x-y$ space in Fig. 1(b). The parameter $\beta$ determines the width of ribbon-like region. In Fig. 1(b), the two dashed blue lines are boundaries of the ribbon-like region, which are mapped from the branch cut (dashed blue line) in Fig. 1(a), with width $L = 2\pi/\beta$. According to Eq. (1), we can derive the refractive index profile in $x-y$ space in Fig. 1(b) as[33]

$$n_z = \frac{n_0}{\cosh(\beta x)}, \tag{2}$$

where $n_0 = \alpha\beta$. Here, if we choose $\alpha = 1$ and $\beta = 1$, the refractive-index profile is that shown on the left of Fig. 1(a). The refractive indices along the dashed purple lines in Fig. 1(a) and Fig. 1(b) are the same. Light rays (red curve with arrows) of Fig. 1(b) can be mapped from those of Fig. 1(a) by $w = e^{\beta z}$, which can also be obtained by the variational method of geometric optics. In fact, we can expand this lens in the $y$ direction to construct a Mikaelian lens, where light rays can be self-focused[32] periodically along the line at x=0, with a half of the periodicity $L$ (Fig. 1(d)). One can also imagine that the whole conformal lens in Fig. 1(d) is mapped from the Riemann surface of exponential conformal mapping shown in Fig. 1(c). The Riemann surface contains infinite numbers of Riemann sheets. Here we only show three of them, each of which is a complex plane endowed with a Maxwell's fish eye lens. They are connected by branch cuts shown as blue dashed lines in Fig. 1(c). Because of the existence of the inhomogeneous lens, a light ray (in red) will travel along a circle. Once it passes through the branch cut, it will go from one Riemann sheet to another. The whole trajectory of a light ray in Riemann surfaces looks like a "spiral" curve. Its conformal image in $x-y$ space is the red sine-like curve in Fig. 1(d). Different Riemann sheets in Fig. 1(c) are mapped to different ribbon-like regions bounded by blue dashed lines in Fig. 1(d).

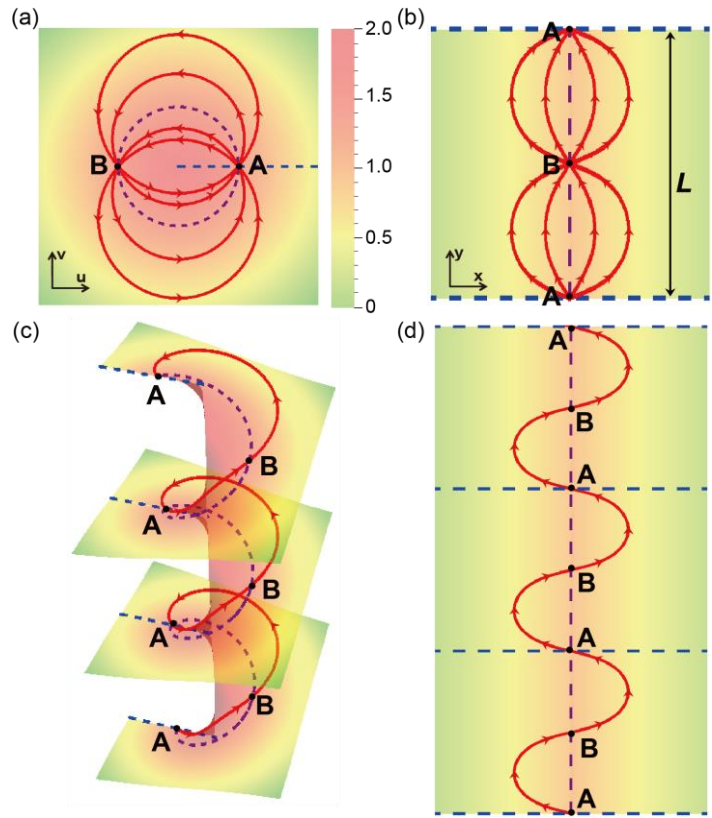

**FIG. 1. Exponential conformal mapping.** (a) Imaging by Maxwell-fish eye in $u-v$ space. (b) Self-focusing of light rays by Mikaelian lens in $x-y$ space. (c) A "spiral" light ray in the Riemann surfaces. (d) A sine-like rays in $x-y$ space.

So far we have constructed a conformal lens (or a Mikaelian lens) in Fig. 1(d) by CTO from a Maxwell's fish-eye lens. Now we will employ a practical experimental system to demonstrate its properties. In the original work of optical conformal mapping, an isotropic medium with a non-uniform refractive index profile was used[4]. Here we propose a new way to visualize CTO by using a gradient-index micro-structured optical waveguide at optical frequencies (Supplemental Material, Section II[50]). As a specific example, we fabricate a Mikaelian lens in a structured waveguide, which is built on an air-PMMA-Ag-SiO$_2$ multilayer structure (Fig. 2(a)). A laser beam of a large spot size is coupled to the waveguide through a grating, and used as a broad incident beam for a Mikaelian lens. Such a beam is denoted in red color in Fig. 2(a) to show its self-focusing property in the geometric-optics limit. In Fig. 2(b), we show effective refractive index of the constructed lens as a red curve (see Section III in Supplemental Material[50]). The black curve is the fitted result with the refractive index profile given by Eq. (2), designed from the conformal mapping shown in Fig.1 (b). Such a lens can be viewed as a Mikaelian lens. As theory proposed in Fig. 1(b), we observed self-focusing in our experiment (see Section IV in Supplemental Material[50]) (Fig. 2(d)). Its focusing length is a quarter of the periodicity $L$. Figure 2(f) depicts a numerical simulation with parameters extracted from the experiment(see Section V in Supplemental Material[50]), the experiment and numerical results are in good agreement with each other. In another experiment (Fig. 2(c)), we use a laser of a small spot size to excite the waveguide and generate a narrow light ray (shown in red). Based on the theory proposed in Fig. 1(d), we also schematically show sine-like rays in the constructed lens. Figs. 2(e) and 2(g) show the experimental and numerical result of light-ray trajectories respectively, which agree very well with the theory. However, there is some absorption in experiment given in Fig. 2(e), mostly caused by fluorescence emission by rare-earth ions excited by the laser beam. This fluorescence emission is just an experiment technique used to measure the light trajectory inside the conformal waveguide (see Supplemental Material Section II for more details[50]), which is not necessary for the device. The absorption can be reduced by removing the rare earth ion in a practical application. In this lens, the width is about 30 $\mu m$, which is about 65 times the working wavelength (0.46 $\mu m$), meaning that the device size is much larger than the wavelength.

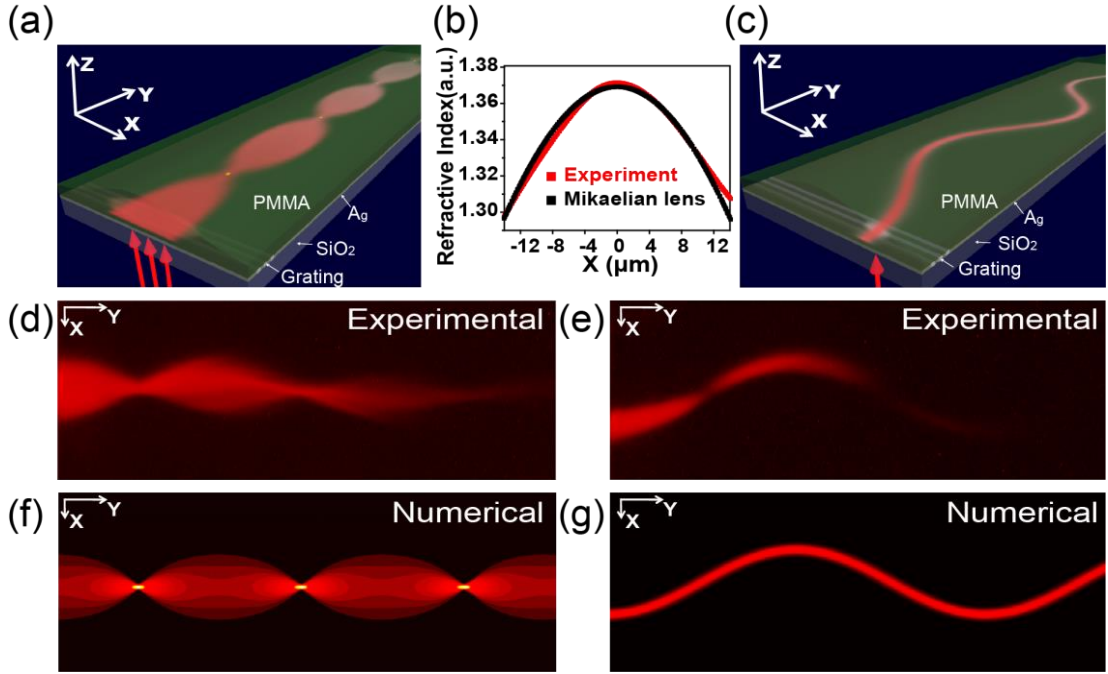

FIG. 2. Schematics and optical measurements of light rays in a conformal waveguide. (a), (c) Schematic view of the micro-structured optical waveguide. (b) The effective refractive index calculated from the waveguide thickness profile (see Supplemental Material Section III for more details[50]). (d), (f), (e), (g) The experiment results and numerical simulations of self-focusing effect (d), (f) and sine-like rays (e), (g) in conformal lens.

The above experiment only demonstrates CTO in the geometric-optics limit from the self-focusing property of a Mikaelian lens in a gradient-index micro-structured optical waveguide. Recently, CTO has also been expanded to wave optics[3,21-31], which enriches its application. In our another experiment, by redesigning the coupling grating in the same waveguide structure we achieve a conformal Talbot effect (Fig. 3(c)). This is an important and interesting effect in wave optics and shows a big difference from the ordinary Talbot effect in a homogeneous medium. For comparison, Figure 3(a) depicts an ordinary Talbot effect with an infinite periodic incident source, where the periodic source pattern repeats along the propagation direction at integer multiples of the primary Talbot length of $2D^2/\lambda$ and is equally spaced along in the transverse direction. In a real practical system, the incident wave cannot be infinitely large. Figure 3(b) shows the results of an ordinary Talbot effect with finite periodic source. It's obvious that, for a finite source, the Talbot effect can only be maintained for a short distance due to boundary effect (see Section VI in Supplemental Material[50]). As a result, a practical ordinary Talbot effect cannot transfer the field pattern without diffraction. However, the conformal Talbot effect in a Mikaelian lens can avoid this diffraction problem (Figs. 3(c)- 3(e)). The replicas of the finite periodic source pattern are vertically squeezed to some special positions at y =L/4, and the input source pattern is perfectly recovered at y =L/2 (Fig. 3(c)). In comparison with the pattern illustrated in Fig. 3(b), it can be seen that the input source pattern can be transferred for a very long distance. In this process there is no diffraction loss, which happens in a finite ordinary Talbot effect (Fig. 3(b)). The numerical simulation is given in Fig. 3(d) and analytical calculation in Fig. 3(e), which agree with the experimental result shown in Fig. 3(c). Details of the analytical solution are provided in Supplemental Material Section VII[50]. In the above two experiments, we observe phenomena characteristic of both geometric optics and wave optics in the same conformal waveguide, demonstrating the capacity of CTO devices.

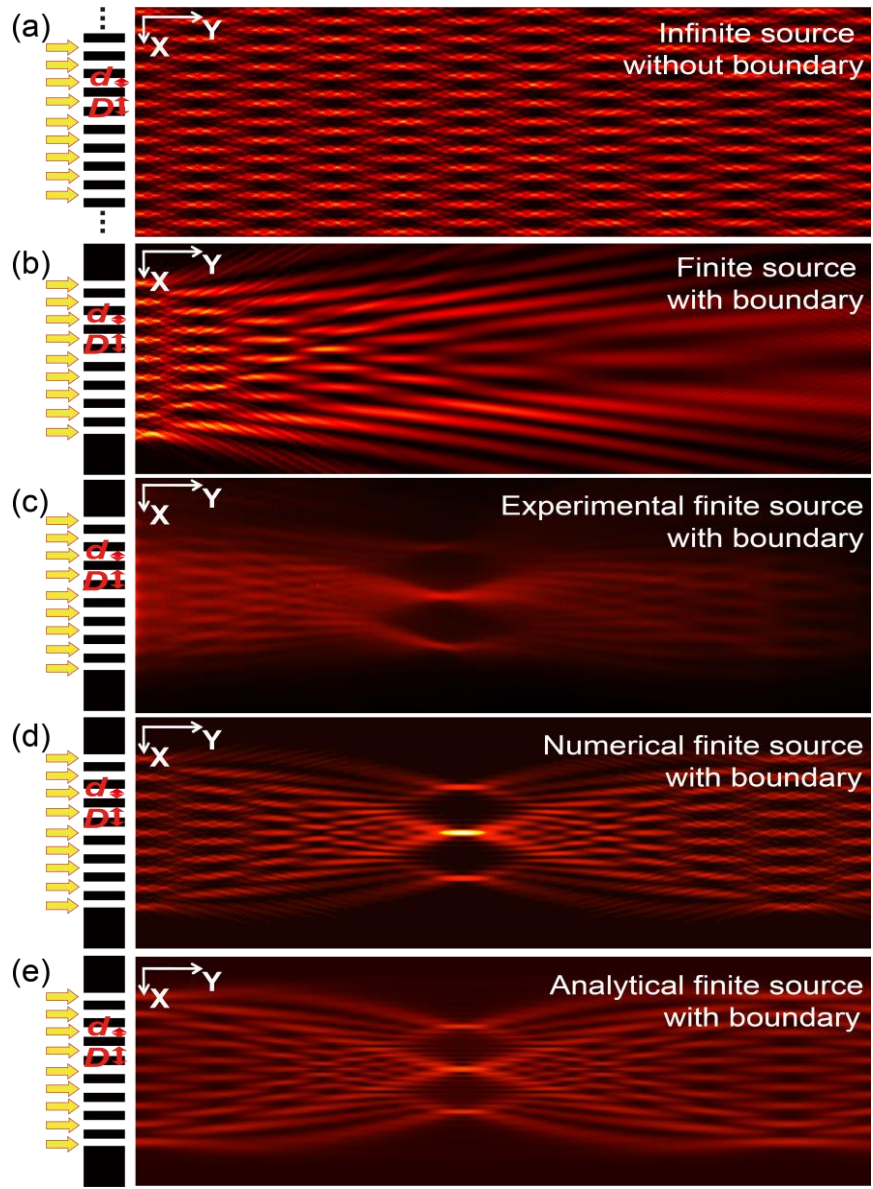

FIG. 3. Comparison between ordinary Talbot effect and conformal Talbot effect. (a), (b) Ordinary Talbot effect in homogeneous medium with infinite periodic source (a) and finite periodic source (b). (c), (d), (e), Self-focusing of the conformal Talbot effect in a Mikaelian lens: (c) experimental results; (d) numerical results; (e) analytical results. The figures in the left column show the schematic of the grating source, the grating period is $D$, the slit width is $d$. An incident laser is used to excite the grating source (denoted as yellow arrows).

It is well-known that information in computer science is restored and transferred as a string of bits: '0' or '1'. Here, we employ the conformal Talbot effect to transfer an encoded field pattern. By tuning the grating parameters D and d, we encode the field pattern with the two bits '0' and '1' (see Supplemental Material Section VIII for more details[50]). Here we use thirty-six bits to investigate the digital coding information transferred through the Talbot effect in a Mikaelian lens waveguide. We find that the coding information can be transferred over a long distance with small distortion. Figures 4(a)-4(c) show that three kinds of coding sources (denoted by Input1, Input2 and Input3) can be focused and transferred in this conformal lens waveguide. Figures 4(d)-4(f) display the coding sources with specific sequences (orange line) and the normalized intensity profiles of the coding field pattern at different propagation distances. The coding sources are represented by the sequences of '0' and '1'. With the coding sources (for example, Input1) imported to this conformal lens waveguide, we calculate the input field pattern at y=0 (red curve in Fig. 4d), focusing pattern at y =L/4 (black curve in Fig. 4(d)) and output field pattern at y =L/2 (blue curve in Fig. 4(d)). It can be clearly seen that the encoded field pattern consists of several separated peaks with specific profiles at the focusing distance, and the magnitudes of the peaks are different for different imported coding sources. Comparing the input and output patterns, we demonstrate that the coding source information could be transferred with very small distortion. Therefore, we can transfer digital coding information efficiently using the conformal Talbot effect, for instance in an optical communication system. One could subsequently transfer this encoded signal into a normal optical chip or waveguide and use another conformal lens at the receiving port to decode the information. Therefore, the conformal Talbot effect has potential application to digital coding transfer without information loss.

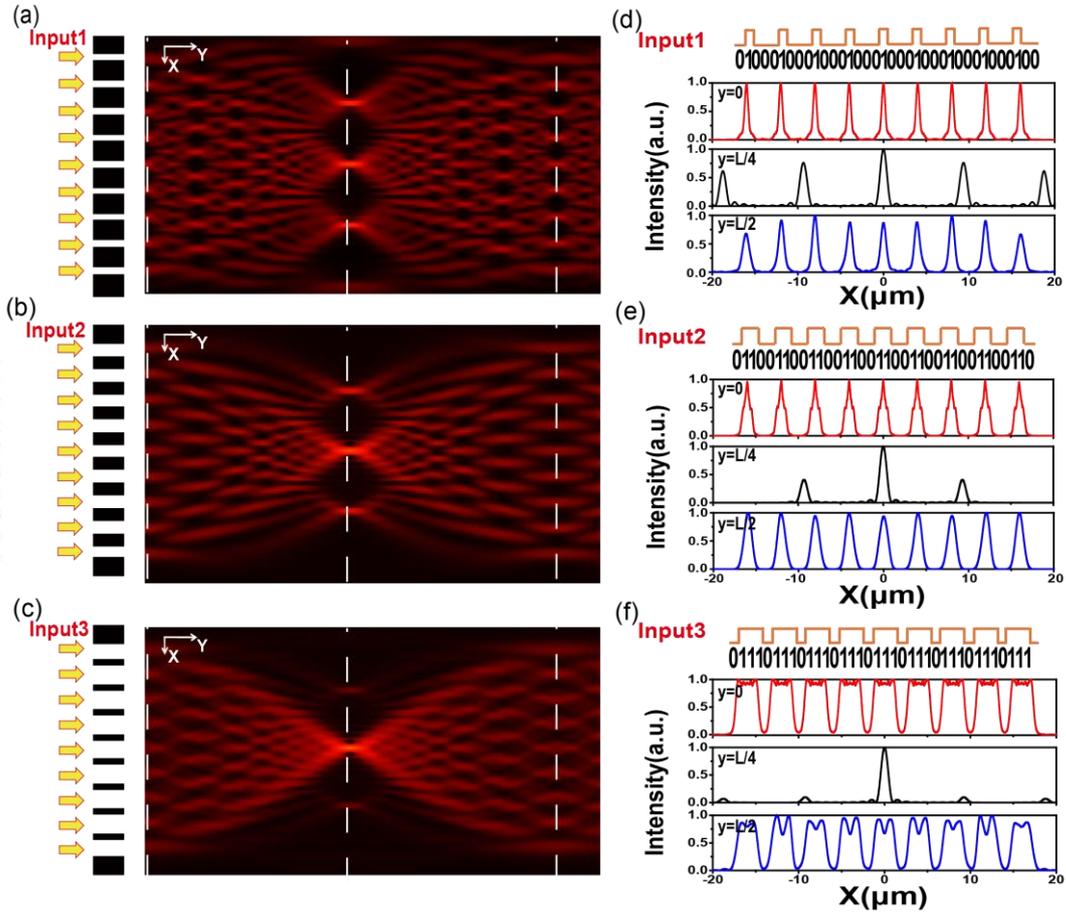

FIG. 4. Digital coding from different source sequences. (a), (b), (c) Field pattern with thirty-six bits coding sequences: (a) 010001000100010001000100010001000100, (b) 011001100110011001100110011001100110, (c) 011101110111011101110111011101110111. The left pictures schematically show the input coding source sequences denoted by Input1, Input2, Input3. (d), (e), (f), The finite periodic coding source sequences(on top) and the normalized intensity profiles for several propagation distances. For y = 0, we get the input signal (red solid curve). The encoded result (black solid curve) is the field at y =L/4, or the focusing plane. For y =L/2, we get the final output result (blue solid curve), which deviates little from the input signal.

In conclusion, we have employed a new platform using a gradient-index micro-structured optical waveguide to realize conformal transformation optics (CTO) devices. Based on conformal mapping theory, we designed and fabricated a conformal lens to simultaneously obtain a self-focusing effect in the geometric-optics limit and a Talbot effect in the wave-optics limit. Numerical simulations and analytical calculations confirm the experimental results, which demonstrate that such a conformal device can work both in geometric optics and wave optics. We also show that this Talbot effect can be used to transfer a digital field pattern without diffraction and has potential applications in digital coding communications.

This work was financially supported by the National Natural Science Foundation of China (No. 11690033, 61322504, 61425018 and 11374151).

# Supplementary Material

# of

# Self-focusing and Talbot effect in conformal transformation optics

Xiangyang Wang[1,a], Huanyang Chen[2,a,b], Hui Liu[1,b], Lin Xu[2], Chong Sheng[1] and Shining Zhu[1]

[1]National Laboratory of Solid State Microstructures & School of Physics, Collaborative Innovation Center of Advanced Microstructures, Nanjing University, Nanjing, Jiangsu 210093, China.

[2]Institute of Electromagnetics and Acoustics and Department of Electronic Science, Xiamen University, Xiamen 361005, China.

[a]These authors contributed equally to this work.

[b] Corresponding author：e-mail: kenyon@xmu.edu.cn; liuhui@nju.edu.cn.

## I. Theory of conformal transformation optics

Conformal transformation optics[4] is based on form invariance of Helmholtz equation under conformal mappings. Suppose that we have two complex planes denoted by $w = u + vi$ and $z = x + yi$, respectively. Let's endow both spaces with refractive index profiles $n_w(u+vi)$ and $n_z(x+yi)$. In $u-v$ space, the Helmholtz equation reads,

$$(\partial_u^2 + \partial_v^2 + n_w^2 k^2)\psi_1 = 4(\partial_w^* \partial_w + n_w^2 k^2)\psi_1 = 0 . \tag{S1}$$

While in $x-y$ space, Helmholtz equation has the same form,

$$(\partial_x^2 + \partial_y^2 + n_z^2 k^2)\psi_2 = 4(\partial_z^* \partial_z + n_z^2 k^2)\psi_2 = 0 . \tag{S2}$$

Considering a conformal mapping $w = w(z)$, which satisfies Cauchy-Riemann condition,

$$\frac{\partial u}{\partial x} = \frac{\partial v}{\partial y} , \quad \frac{\partial u}{\partial y} = -\frac{\partial v}{\partial x} , \tag{S3}$$

then $\partial_w^* \partial_w = |dz/dw|^2 \partial_z^* \partial_z$. Equation (S1) changes into

$$4(\partial_z^* \partial_z + (n_w |dw/dz|)^2 k^2)\psi_1 = 0.  \tag{S4}$$

Comparing Eq. (S2) and Eq. (S4), we know that $\psi_2(z) = \psi_1(w(z))$, if

$$n_z = \left|\frac{dw}{dz}\right| n_w. \tag{S5}$$

We have explained the physics of wave propagation in $u-v$ space and $x-y$ space are equivalent, if these two spaces are connected by a conformal mapping $w = w(z)$ and their refractive index profiles satisfy Eq. (S5). For simplicity, people usually call $u-v$ space as virtual space, and $x-y$ space as physical space.

## II. Sample fabrication

A micro-structured waveguide with refractive index profile of Mikaelian lens was fabricated for our experiment, as depicted in Fig. 2(a) of our manuscript. During the sample fabrication process, a 50 nm thickness silver film was sputtered onto the silica substrate, followed by a straight gold cylinder (diameter = $100\ \mu m$) glued to the substrate. Next, a coupling grating with a period of 310 nm was milled on the silver film at the side of the cylinder with focused ion beam (FEI Strata FIB 201, 30 keV, 11 pA). These gratings were used to couple laser into the waveguide. Next, we spin-coated the sample with polymethyl methacrylate (PMMA) resist, and subsequently dried the sample in the oven at 70°C for 2 h. Due to the surface intension of PMMA solution and the gold cylinder, an inhomogeneous PMMA waveguide was formed near the cylinder with changed thickness. The thickness profiles can be changed by varying the solubility of the PMMA solution, evaporation rate and spin rate. In our samples, the waveguide thicknesses remain uniform in the direction parallel to the axis of the cylinder in y direction, while it shows continuous change along the x direction. The schematic view of the waveguide is shown in Figs. 2(a) and 2(c) of our manuscript, where we can see that the micro-structured waveguide consists of an air/PMMA/Ag/SiO2 multilayer stack and can be considered as a step-index planar waveguide. Through the coupling grating, we can excite one mode for 460 nm laser light inside the waveguide. The rare earth ions $Eu^{3+}$ was added to the PMMA resist to facilitate fluorescence imaging that would reveal the propagation dynamics of the beam. The $Eu^{3+}$ absorbed the light (at 460nm wavelength) propagating in the waveguide, and then emitted fluorescent light at 615 nm. Here, only $TE_0$ mode was excited in our experiment.

### III. Experimental measurements, fittings and effective refractive index

The polymethyl methacrylate (PMMA) waveguide thickness profile was directly measured using surface profilometer (Veeco DEKTAK 150 Profilometer), which can be also directly measured by the interference pattern obtained in the transmission image of white light through the sample. The results are shown in Fig. S1. The interference pattern (Fig. S1(a)) shows the waveguide thickness is uniform in the y direction, while symmetrically changes along the x direction. Figure S1(b) depicts the waveguide thickness profile in the x direction. The measured thickness $t$ can be fitted with a polynomial function of the type,

$$t(x) = a_0 + a_1 x + a_2 x^2 + a_3 x^3 + a_4 x^4 + a_5 x^5 ,  \tag{S6}$$

where $a_0$, $a_1$, $a_2$, $a_3$, $a_4$ and $a_5$ are constants. The thickness fitting parameters are included in Table S1. The effective refractive index of PMMA waveguide is calculated by considering the micro-structured waveguide geometry consisting of Air/PMMA/Ag/SiO2 multilayer stack. The thickness of metal film (Ag) is 50 nm. Clearly, the Ag layer is thick enough to reduce the problem into a single dielectric waveguide on a top of a semi-infinite metal substrate. For transverse electric (TE) waves, the dispersion relationship is given as:

$$k_1 t = \arctan(k_2/k_1) + \arctan(k_3/k_1) + m\pi \quad (S7)$$

where $t$ is the PMMA layer thickness, $m$ is the mode index, and $k_1 = (2\pi/\lambda)\sqrt{n_1^2 - n_{eff}^2}$, $k_2 = (2\pi/\lambda)\sqrt{n_{eff}^2 - n_2^2}$, $k_3 = (2\pi/\lambda)\sqrt{n_{eff}^2 - n_3^2}$ are the transversal wave vector in the PMMA layer ($n_1$), Air($n_2$), Ag substrate($n_3$), correspondingly. $n_{eff}$ is the effective waveguide refractive index.

The dispersion relationship for TE modes are calculated using Eq. (S7). In the calculation we use: $n_1$ = 1.49, $n_2$ = 1, and the Ag substrate[51] $n_3$ = $0.13576 + 2.46806i$ at the exciting wavelength $\lambda = 460$ nm. The estimated effective refractive index for TE$_0$, the mode index ($m = 0$ mode), is shown in Fig. S1(c) (red solid line). The refractive index is then fitted to a Mikaelian lens refractive index function of the type,

$$n(x) = \frac{n_0}{\cosh(\beta x)}, \quad (S8)$$

where $\beta$ determines the width of Mikaelian lens. The fitting parameters are included in Table S2.

**TABLE S1** Experimental fitting parameters for the polymethylmethacrylate (PMMA) film thickness

| Data type | $a_0[nm]$ | $a_1[nm/\mu m]$ | $a_2[nm/\mu m^2]$ | $a_3[nm/\mu m^3]$ | $a_4[nm/\mu m^4]$ | $a_5[nm/\mu m^5]$ |
|---|---|---|---|---|---|---|
| Fitting results | 272.49159 | 0.90761 | −0.69346 | −0.00489 | 0.00157 | $1.19739*10^{-5}$ |

**TABLE S2** Experimental fitting parameters for the effective refractive index profile

| Data type | $n_0$ | $\beta [\mu m^{-1}]$ |
|---|---|---|
| Fitting results | 1.36918 | 0.02374 |

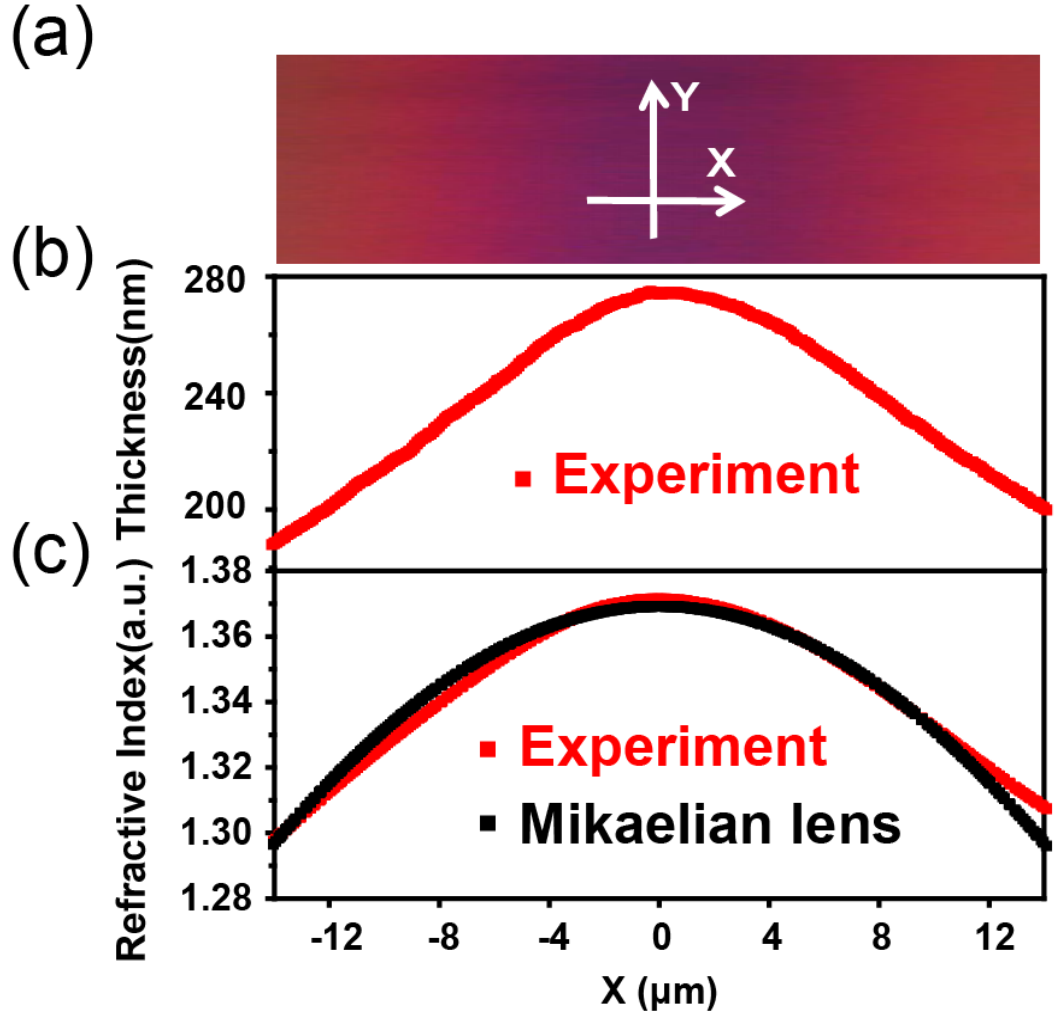

**FIG. S1. (a)**, Interference pattern in the micro-structured waveguide illuminated by white light. **(b)**, Film thickness measured using surface profilometer. **(c)**, The effective refractive index calculated using Eq. (S7) (red solid curve) and the Mikaelian lens refractive index(black solid curve).

### IV. Optical measurements

For measurement of the light propagation in a Mikaelian lens waveguide, we coupled a coherent continuous-wave laser with a wavelength of $\lambda = 0.46 \ \mu m$ into the waveguide by a grating (shown in Figs. 2(a) and 2(c) of our manuscript). When the coupled light propagates in the waveguide, it would stimulate the rare earth ions, which will then re-emit fluorescence light at another wavelength of 615 nm. The fluorescence light from the rare earth ions is collected by a microscope objective and delivered to a charge-coupled device camera. In this way, the light trajectories and optical field pattern inside the conformal waveguides can be observed directly.

### V. Numerical calculations

For the numerical simulations we employed the commercial FDTD software (Lumerical Solutions, Inc.). In all simulations the parameters used are extracted from the experimental data. In the main text, Figures 2(f), 2(g) and Figures 3(a), 3(b), 3(d) of our manuscript are numerical results from the FDTD software.

### VI. The boundary effect on Talbot effect

In this part, we study the boundary effect on Talbot effect with a finite periodical source- that is, Talbot effect in a homogeneous medium and in a Mikaelian lens waveguide. To address this, we theoretically analysis the wave interference of finite periodical source with different length in a homogeneous and a Mikaelian lens waveguide, respectively.

Figure S2(a) depicts Talbot effect for different sources in a homogeneous waveguide, while Figure S2(b) shows Talbot effect for the same source in a Mikaelian lens waveguide. We observe enormous difference of light propagating between the homogeneous and Mikaelian lens waveguide. In the homogeneous waveguide, the wave interference is becoming weaker and weaker with the increase of propagation distance, thereby the periodical source pattern cannot be reproduced. This is due to the boundary effect in the homogeneous waveguide. But in the Mikaelian lens waveguide, we see that the periodical source pattern can be transferred for a very long distance with very small distortion. This is due to the fact that we put an inverse hyperbolic cosine refractive index profile in the transverse direction.

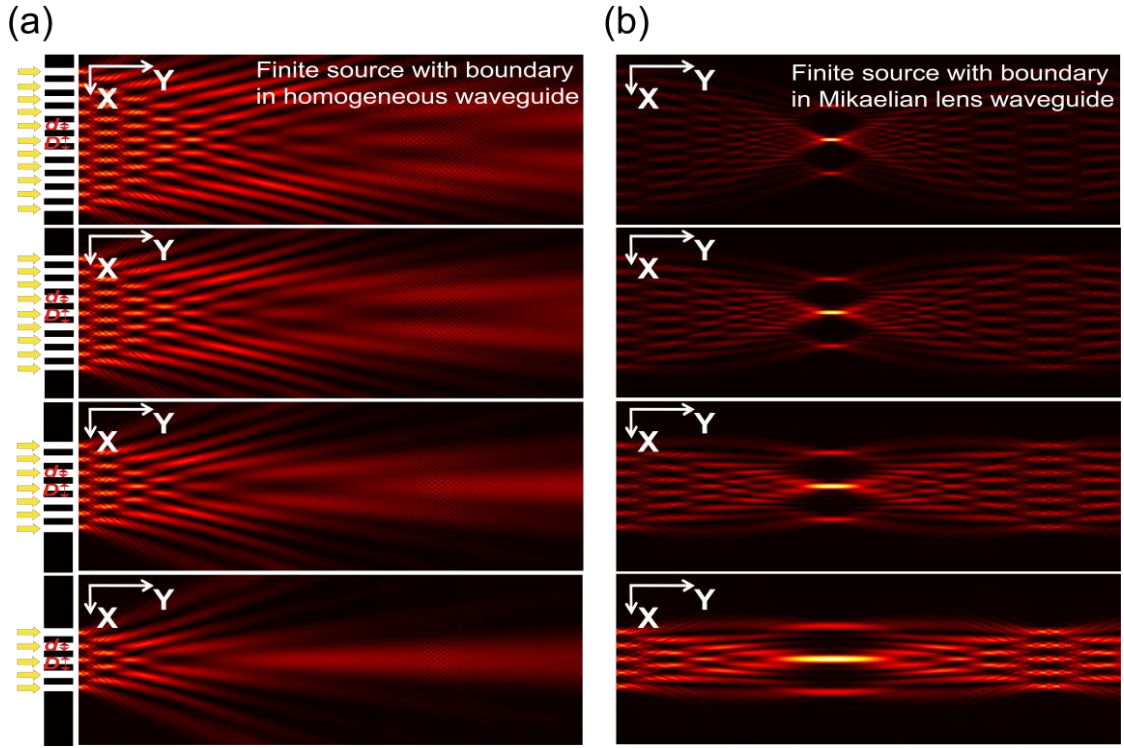

**FIG. S2.** Boundary effect on Talbot effect. **(a),** Talbot effect in a homogeneous waveguide. **(b),** Talbot effect in a Mikaelian lens waveguide. From top to bottom, the length of the periodic source is gradually decreased: 11 periods, 9 periods, 7 periods and 5 periods. The period $D$ and slit width $d$ respectively are $4~\mu m$ and $2~\mu m$.

### VII. Analytical solution for Talbot effect in Mikaelian lens waveguide

The Mikaelian lens planar waveguide is a two dimensional gradient medium whose refractive index is given by Eq. (S8). The Helmholtz equation for the electric vector projection of the TE wave, $E_z(x, y)$, can be given as,

$$\left[\frac{\partial^2}{\partial x^2}+\frac{\partial^2}{\partial y^2}+\frac{k_0^2 n_0^2}{\cosh^2(\beta x)}\right]E_z(x,y)=0, \tag{S9}$$

where $k_0=2\pi/\lambda$ is the vacuum wavenumber and $\lambda$ is the wavelength in free space. Here, the field of Talbot effect in Mikaelian lens can be decomposed into modes described by the Jacobi polynomials[52,53]. The mode solution for Eq. (S9) takes the following form,

$$E_z^m(x,y)=\exp(i\beta\gamma y)(1-\eta^2)^{\gamma/2}P_m^{(\gamma,\gamma)}(\eta), \tag{S10}$$

where $\gamma=\left\{\left[1+4(k_0 n_0/\beta)^2\right]^{1/2}-(2m+1)\right\}/2$, $\gamma$ is the propagation constant of the mode, $\eta=\tanh(\beta x)$, $m$ is a nonnegative integer, $P_m^{(\gamma,\gamma)}$ are the normalized Jacobi polynomials. An arbitrary solution of Eq. (S9) can then be decomposed in terms of the basis in Eq. (S10):

$$E_z(x,y)=\exp(i\beta\gamma_0 y)\sum_{m=0}^{\infty}C_m\exp(-i\beta m y)\psi_m(x), \tag{S11}$$

where $\gamma_0=\gamma+m$, $\psi_m(x)=(1-\eta^2)^{\gamma/2}P_m^{(\gamma,\gamma)}(\eta)$, $C_m$ is coefficient of the m-th mode, and $C_m$ can be represented as,

$$C_m=\int_{-\infty}^{+\infty}\psi_m^*(x)E_z(x,0)dx \tag{S12}$$

where $E_z(x,0)$ is the input electric field at the plane of y = 0. The input field (Fig. S3(a)) is a finite periodic source. The corresponding coupling grating in experiment is shown in Fig. S3(b), while the field amplitude is given in Fig. S3(a). Using the parameters from experiments, we can calculate the coefficient $C_m$, given in Table S3. The corresponding analytical result (Fig. 3(e) in our manuscript) agrees well with the experimental and numerical result (Figs. 3(c) and 3(d) in our manuscript, respectively).

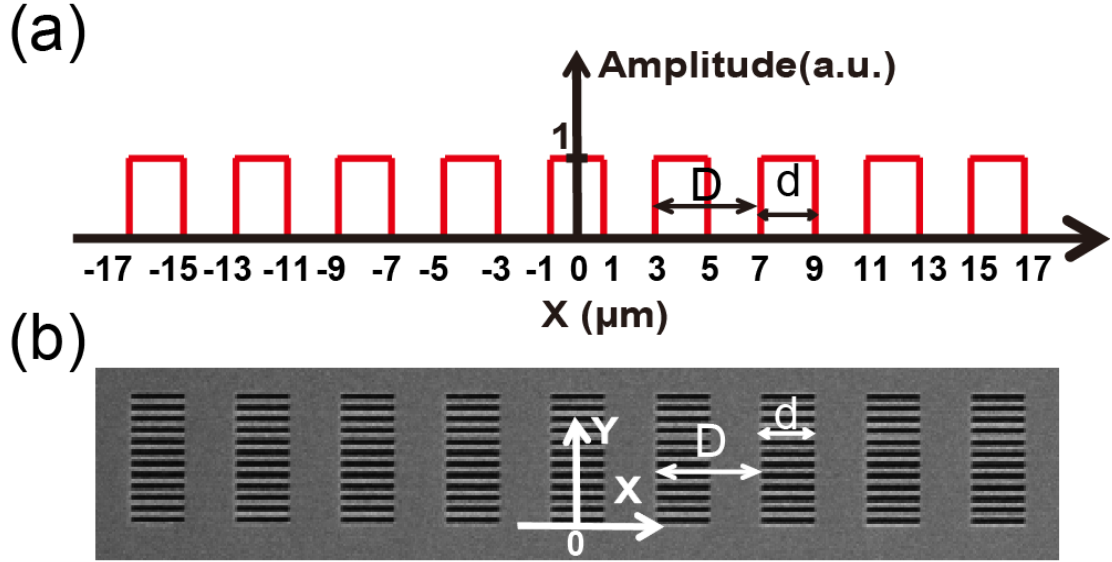

**FIG. S3. (a),** The input field amplitude at $y=0$, with a period $D=4\ \mu m$ and a slit width $d=2\ \mu m$. **(b),** SEM picture of the coupling grating.

**TABLE S3** Mode coefficient fitted based on Eq. (S12)

| $C_m$ | $C_0$ | $C_2$ | $C_4$ | $C_6$ | $C_8$ | $C_{10}$ | $C_{12}$ | $C_{14}$ | $C_{16}$ | $C_{18}$ | $C_{20}$ | $C_{22}$ | $C_{24}$ | $C_{26}$ | $C_{28}$ | $C_{30}$ |
|---|---|---|---|---|---|---|---|---|---|---|---|---|---|---|---|---|
| **Fitting results** | 0.7748 | 0.5595 | 0.3973 | 0.7239 | -0.1909 | 0.9730 | 0.4022 | -0.1573 | 0.3552 | 0.7932 | 0.4753 | -0.0211 | -0.0396 | 0.3786 | 0.651 | 0.5906 |

**VIII. Digital coding**

In this section, we will discuss the digital coding in Mikaelian lens. To realize the digital coding to mimic the binary concept in computer science, we introduce two primary unit cells: "0" and "1" elements (Fig. S4) for 1-bit. "0" denotes an input source with length = 1 μm and

amplitude = 0, while "1" is expressed as an input source with length = 1 μm and amplitude = 1, illustrated in Fig. S3. Here, we take thirty-six bits to investigate the digital coding. Then the finite periodic source in Fig. S3 can be reformulated into a coding sequence of 011001100110011001100110011001100110. By designing coding sequences of '0' and '1' elements in source, we can transfer the source information to different profiles at the focusing plane of the conformal lens (Fig. 4 in our manuscript).

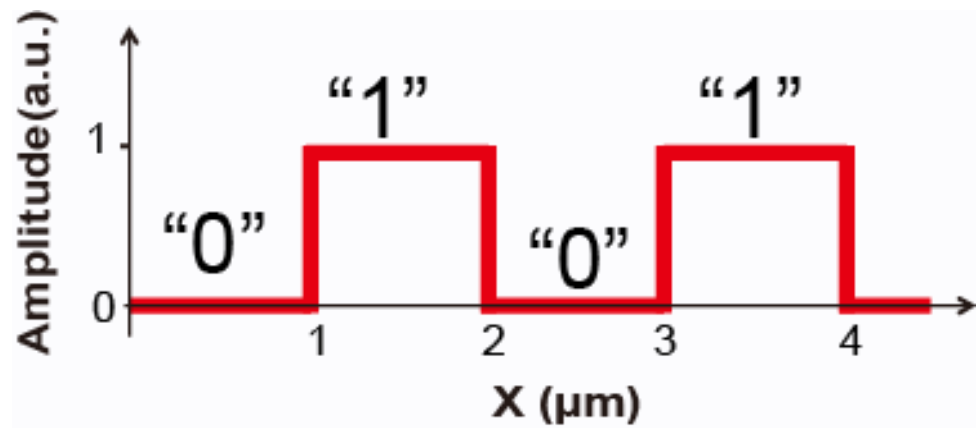

**FIG. S4.** Two type unit cells: "0" and "1" elements for 1-bit.